\documentclass[pre,showpacs,twocolumn,amssymb,bibnotes]{revtex4}
\usepackage{here}
\usepackage[dvips]{graphicx}

\newcounter{Fig}

\begin{document}

\title{Bandgap engineering and defect modes
in photonic crystals \\ with rotated hexagonal holes}

\author{Aaron Matthews$^1$}
\author{Sergei F. Mingaleev$^{1,2}$}
\author{Yuri S. Kivshar$^1$}

\affiliation{$^1$Nonlinear Physics Group and Center for Ultra-high
bandwidth Devices for Optical Systems (CUDOS), Research School of
Physical Sciences and Engineering, Australian National University, Canberra ACT 0200, Australia\\
$^2$Institut f{\"u}r Theorie der Kondensierten Materie,
Universit{\"a}t Karlsruhe, 76128 Karlsruhe, Germany}

\begin{abstract}
We study the bandgap structure of two-dimensional photonic
crystals created by a triangular lattice of rotated hexagonal
holes, and explore the effects of the reduced symmetry in the
unit-cell geometry on the value of the absolute bandgap and the
frequencies of localized defect modes. We reveal that a maximum
absolute bandgap for this structure is achieved for an
intermediate rotation angle of the holes. This angle depends on
the radius of the holes and the refractive index of the background
material. We also study the properties of the defect modes created
by missing holes, and discuss the mode tunability in such
structures.
\end{abstract}


\maketitle

During recent years we observe a rapidly growing interest in the
design and fabrication of novel types of photonic-crystal
structures possessing large absolute band gaps. The study of
various geometries of three-dimensional photonic crystals and
different ways to enlarge their absolute bandgaps is a key issue
of the physics of periodic dielectric structures where different
polarizations of light are coupled together and, therefore, the
existence of an absolute bandgap is crucially important for
various applications of three-dimensional photonic crystals in
optics (see. e.g., Ref.~\cite{sakoda} and references therein).

However, for two-dimensional (2D) photonic crystals Maxwell's
equations are known to decouple effectively for two polarizations,
so that the study of a particular photonic bandgap of such
periodic structures can be carried out independently for both
$E$-polarized and $H$-polarized electromagnetic waves.
Accordingly,  two types of photonic bandgaps are usually
distinguished to exist for different polarizations. When the
bandgaps for two different polarizations overlap, they create the
combined bandgap known as {\em an absolute bandgap}.

One of the main reasons to enlarge the absolute bandgaps and to
study the bandgap properties of 2D periodic photonic structures is
an attractive possibility of creating novel types of tunable
waveguides and circuits for applications of photonic crystals in
integrated optics. In particular, the existence of a large
absolute bandgap would allow to design the waveguides in planar
structures which can support propagating modes of both
polarizations in the same frequency domain.

Up to now, a number of different approaches has been suggested for
the enlargement of the absolute bandgaps of 2D photonic-crystal
structures. One of these approaches relies on the idea of using
the photonic crystals created by a two-dimensional lattice of
non-circular holes and exploring the symmetry-reducing properties
of 2D photonic crystals for the bandgap
enlargement~\cite{Villeneuve:1992-4969:PRB}--\cite{PRB_2003}. In
particular, Wang {\em et al.}~\cite{Wang:2001-4307:JAP}
demonstrated that the absolute bandgap becomes maximal in the case
of air holes of the same symmetry as the lattice symmetry. For
example, the bandgap takes a maximum value for a triangular
lattice of rotated hexagonal holes. However, Qui {\em et
al.}~\cite{Qiu:2000-1027:JOSB} suggested that the bandgap as large
as that demonstrated in Ref. \cite{Wang:2001-4307:JAP} can be
achieved in 2D photonic crystals created by air holes of more
complex, non-circular shape without any rotation.

In spite of many studies of the absolute bandgaps of photonic
crystals created by a lattice of non-circular holes, no specific
applications of such large bandgaps have been discussed and
demonstrated. In particular, the usual statement that large
absolute bandgaps should be useful for functioning of photonic
circuits is not well-grounded. As a matter of fact, there is no
solid reason to exploit the absolute bandgaps in two-dimensional
structures unless photonic-crystal circuits guide light of both
polarizations, but the study of defect modes and waveguides in
such structures is still incomplete or missing.

The purpose of this paper is twofold.  First, we explore further
the concept of the enlargement of the absolute bandgaps of 2D
photonic crystals through reducing symmetry in the unit-cell
geometry taking as an example a 2D dielectric periodic structure
created by a  a triangular lattice of rotated hexagonal holes. In
particular, we reveal that in such 2D structures large absolute
bandgaps can be achieved for an intermediate value of the rotation
angle of the hexagonal holes. Second, we demonstrate that, by
using 2D photonic crystals with a large absolute bandgap, it is
possible to create defects which can support localized modes for
both polarizations of light, so that the waveguides based on such
defects can guide light of both polarizations as well.

\begin{figure}
\centerline{\includegraphics[clip,width=75mm]{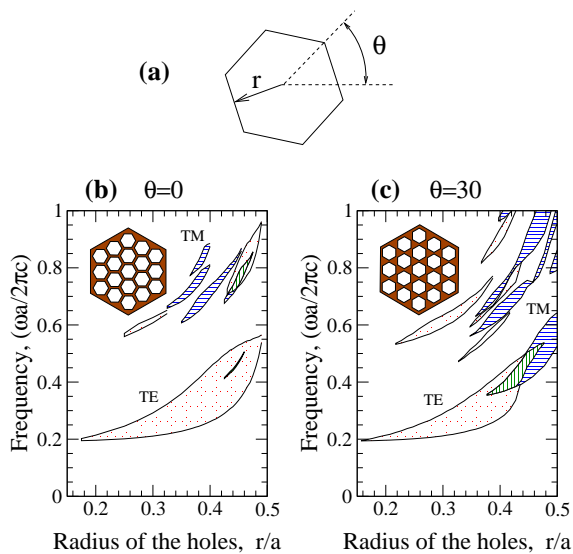}}
\caption{ \label{gaps} Photonic bandgaps of a triangular lattice
of hexagonal holes with the radius $r$, rotated by the angle
$\theta$ [see (a)], in the dielectric with $\epsilon =12$ (silicon
or AsGa) for (b) conventionally oriented hexagonal holes ($\theta
=0^o$), and (c) hexagonal holes rotated by the angle
$\theta=30^o$. Shown are the bandgaps for the both TE (dotted
areas) and TM (horizontally hatched areas) polarized waves,
respectively. Absolute bandgap (hatched vertically) appears as an
overlap of two bandgap structures.}
\end{figure}

We consider a 2D photonic crystal created by a triangular lattice
of hexagonal holes assuming an arbitrary rotation of the hole
relative to the lattice symmetry axis, as shown in
Fig.~\ref{gaps}(a). The photonic band gap is calculated by solving
Maxwell's equations, by means of the plane-wave expansion
method~\cite{PW_1}. Two examples of the photonic bandgap structure
of such a photonic crystal is shown in Figs.~\ref{gaps}(b,c), for
the particular cases of the hole rotation, i.e. for $\theta=0$ and
$\theta=30^o$, respectively.  Our main goal here is to study the
effect of the hole rotation on the value of the (lowest-order)
absolute band gap. The results can  naturally be compared with the
bandgap spectra of the 2D structures created by circular holes, as
well as with the case when the hexagonal holes are not rotated,
see Fig.~\ref{gaps}(b).

\begin{figure}
\centerline{\includegraphics[clip,width=70mm]{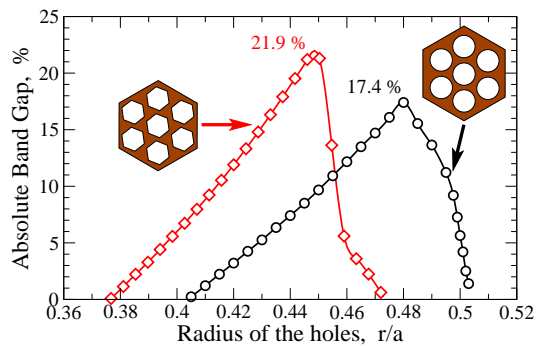}}
\caption{ \label{compar} Comparison between the absolute bandgaps
of two-dimensional photonic crystals created by a lattice of
cylindrical (circles) and rotated hexagonal (diamonds) holes for
different values of the hole radius $r$. For the hexagonal holes
the rotation is chosen to maximize the absolute bandgap at a given
radius.}
\end{figure}

In Fig.~\ref{compar} we show the dependence of the normalized
width of the absolute bandgap for two types of triangular lattices
created by circular and hexagonal holes, as the functions of the
normalized radius, $r/a$. We make this type of comparison for
different values of the rotation angle of the hexagonal lattices,
such that the rotation for the hexagonal holes is chosen to
maximize the absolute bandgap at a given radius. In all the cases
we confirm that the hexagonal lattices possess a larger absolute
bandgap, and this bandgap depends strongly on the hexagon
orientation.

\begin{figure}
\centerline{\includegraphics[clip,width=65mm]{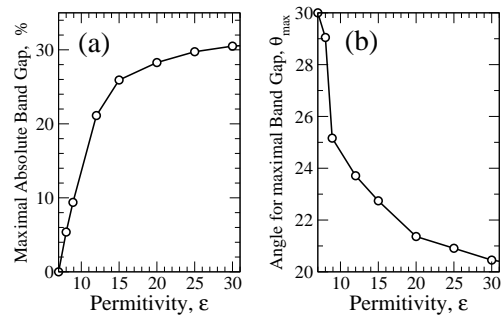}}
\caption{ \label{angle} (a) The maximum value of the absolute
bandgap $\Delta \omega/\omega$ and (b) the corresponding critical
rotation angle $\theta_{\rm cr}$ as functions of the dielectric
permittivity $\epsilon$ of the photonic crystal material.}
\end{figure}

As a matter of fact, we observe that the absolute bandgap can be
enlarged dramatically by employing the concept of local symmetry
reduction in the unit cell of the structure, and using hexagonal
holes instead of circular holes to reduce the rotational symmetry.
In this latter case, the bandgap depends strongly on the rotation
angle of the holes. In brief, our findings can be summarized as
follows. First, the maximal bandgap is achieved for some
intermediate (critical) angle of rotation; in the case of the
hexagonal holes in the material with $\epsilon=12$, this critical
angle is close to the value 24$^o$. Second, the critical angle
$\theta_{\rm cr}$ depends on the value of the dielectric constant
$\epsilon$ of the photonic-crystal material, but it varies slowly
near this intermediate value. Figures~\ref{angle}(a,b) show the
maximum value of the absolute bandgap $\Delta \omega/\omega$ and
the variation of the critical angle $\theta_{\rm cr}$ of the
rotated hexagonal holes on the value of the dielectric
permittivity $\epsilon$ of the photonic crystal material. It is
therefore clear that this critical rotation angle is not a
fundamental constant of the structure, but it is defined by some
effective geometry and material properties corresponding to the
maximum effect of the local symmetry reducing in the unit-cell
geometry, similar to the effects discussed in Ref.
\cite{Qiu:2000-1027:JOSB} for a complete different problem.
Moreover, the absolute bandgap varies dramatically for relatively
small values of $\epsilon$, whereas it saturates for large
$\epsilon$, and the critical angle approaches the value $21^o$.



As we mentioned above, in the case of 2D periodic dielectric
structures, both $E$- and $H$-polarization components of the
electromagnetic field are decoupled. In this case, the existence
of large absolute bandgaps can be useful to support localized
modes for both polarizations of light, and that the waveguides
based on such defects can guide light of both polarizations as
well. Therefore, in order to employ some practical advantages of
using the properties of the absolute bandgaps in 2D photonic
structures, we should explore the properties of the defect modes
and waveguides created in such types of photonic crystals. To be
practically useful, such waveguides should be tolerant to the
fabrication disorder, so that two-dimensional photonic structures
created by a lattice of holes seem to be the best suited
structures where defect modes and waveguides can be fabricated
making some missing holes.

\begin{figure}
\centerline{\includegraphics[clip,width=70mm]{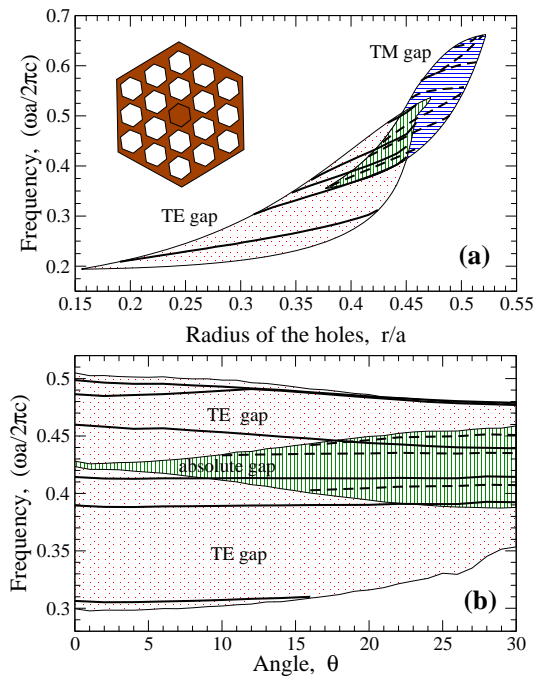}}
\caption{ \label{defect} Spectrum of localized defect modes in a
triangular lattice of rotated hexagonal holes (a) as a function of
the normalized hole radius $r/a$ at the fixed value of the
rotation, $\theta=23.7^o$, and (b) as a function of the rotation
angle $\theta$ for the fixed value of the hole radius, $r=0.43a$.
The defect is created by non-etching a single hole, as shown in
the insert.}
\end{figure}

Figure~\ref{defect}(a) shows the spectrum of localized defect
modes supported by a triangular lattice of hexagonal holes, for
the case when the lattice is created by the holes rotated by
23.7$^o$, and the bandgap takes its maximum value. As follows from
these results, the defect created by a missing rotated hole can
support simultaneously localized modes of both TE and TM
polarizations, allowing for an effective manipulation of light in
photonic-crystal circuits based on this type of defects.
Additionally, in Fig.~\ref{defect}(b) we show how the absolute
bandgap and the defect frequencies vary with a change of the
rotation angle $\theta$, at the fixed value of the hole radius
$r/a=0.43$ defined in Fig.~\ref{gaps}(a).  From the results
presented in Fig.~\ref{defect}(b) it follows that the frequencies
of the defect modes inside the absolute bandgap are almost
constant for all angles.

Our results show that the similar features are observed for the
waveguides created in the lower-symmetry photonic crystals
possessing a large absolute photonic bandgap: the possibility to
enlarge the bandgap in a 2D photonic structures created by rotated
hexagonal holes allows such waveguides to support the guided modes
of both polarization, and their bandgap properties can be easily
manipulated. This problem will be addressed in our future
publications.

In conclusion, we have presented the results of engineering of the
absolute bandgaps in 2D photonic crystals for photonic-circuit
applications, for the example of a photonic crystal created by a
triangular lattice of hexagonal holes. We have revealed that the
maximum value of the absolute bandgap in this structure can be
achieved when all hexagonal holes are rotated by {\em a finite
angle}, and this rotation angle takes an intermediate value
between the values corresponding to the simplest symmetries of the
lattice. The critical rotation angle depends on the parameters of
photonic crystals, but it is shown to be almost constant for large
values of the dielectric permittivity. Moreover, the frequencies
of the defect modes which can be supported by missing holes in
such structures do not vary much with the hole rotation and
changing the value of the absolute bandgap.

We believe that our results will be important for a design of the
photonic-crystal waveguides and circuits supporting propagating
guided modes of both polarizations in the same frequency domain,
as well as for the study of nonlinear waveguides where the
intensity-induced coupling between the modes of different
polarization can be controlled through the rotation of holes or
non-circular defects. In addition, we believe that our results can
be useful for a design of novel types of coupled-resonator optical
waveguides with the properties which can be tuned by using rotated
hexagonal holes instead of circular holes.

In addition, we would like to mention that large absolute bandgaps
can be very useful for exploring nonlinear properties of photonic
crystals, when the intensity-dependent refractive index of the
waveguides can be employed to couple the field polarizations. In
this case, we can create nonlinear photonic-crystal circuits where
one polarization is used for the signal transmission, whereas the
other polarization is employed for controlling the signal
propagation, in order to realize switching between different
transmission regimes, etc.

This work has been supported by the Australian Research Council
through the ARC Center of Excellence program and the Center for
Ultra-high bandwidth Devices for Optical Systems (CUDOS). We
thanks Dr. X.H. Wang for useful discussions.

\end{document}